\documentclass[10pt,journal,compsoc]{IEEEtran}

\usepackage{booktabs}
\usepackage{tabularx}
\usepackage{array}
\usepackage{multirow}
\usepackage{graphicx}
\usepackage{amsmath}
\usepackage{amssymb}
\usepackage{url}
\usepackage{xurl}
\usepackage{xcolor}
\usepackage{listings}
\usepackage{enumitem}
\usepackage[hidelinks,hypertexnames=false]{hyperref}
\usepackage{tikz}
\usetikzlibrary{positioning,shapes.geometric,shapes.symbols,arrows.meta,fit,backgrounds,calc,decorations.pathreplacing,patterns}
\usepackage{fontawesome5}
\usepackage{algorithm}
\usepackage{algpseudocode}
\algrenewcommand\algorithmicrequire{\textbf{Input:}}
\algrenewcommand\algorithmicensure{\textbf{Output:}}
\algrenewcommand\textproc{\textsc}

\definecolor{telcolor}{HTML}{2C6FB1}    % telemetry / read: blue
\definecolor{graphcolor}{HTML}{5C9479}  % causal graph / source: green
\definecolor{agentcolor}{HTML}{C8721A}  % planner / agentic reasoning: amber
\definecolor{execcolor}{HTML}{36206B}   % guarded execution / output: indigo
\definecolor{feedbackcolor}{HTML}{B03A48} % feedback / safety: coral

\tikzset{
  iconlabel/.style={font=\small\bfseries},
  stepnum/.style={
    circle, draw=black!70, fill=white, line width=0.5pt,
    inner sep=0pt, minimum size=3.6mm,
    font=\scriptsize\bfseries
  },
}

\newcommand{\circled}[1]{\tikz[baseline=(c.base)]{\node[circle,draw=black!70,line width=0.4pt,inner sep=0.4pt,minimum size=2.6mm,font=\tiny\bfseries](c){#1};}}

\lstdefinelanguage{json}{
  basicstyle=\ttfamily\footnotesize,
  showstringspaces=false,
  breaklines=true,
  frame=single,
  columns=fullflexible,
  string=[s]{"}{"},
  comment=[l]{:},
  morecomment=[l]{//},
  literate=
   *{0}{{{\color{black}0}}}{1}
    {1}{{{\color{black}1}}}{1}
    {2}{{{\color{black}2}}}{1}
    {3}{{{\color{black}3}}}{1}
    {4}{{{\color{black}4}}}{1}
    {5}{{{\color{black}5}}}{1}
    {6}{{{\color{black}6}}}{1}
    {7}{{{\color{black}7}}}{1}
    {8}{{{\color{black}8}}}{1}
    {9}{{{\color{black}9}}}{1}
}

\lstdefinestyle{algo}{
  basicstyle=\ttfamily\footnotesize,
  showstringspaces=false,
  breaklines=true,
  frame=single,
  columns=fullflexible
}

\newcommand{\sys}{ARBITER}
\newcommand{\otel}{OpenTelemetry}

\title{\sys: Guarded Agentic Control for SLO-Oriented Kubernetes Remediation}

\author{Pooyan~Habibi,~\IEEEmembership{Member,~IEEE,}
        and~Alberto~Leon-Garcia,~\IEEEmembership{Life~Fellow,~IEEE}%
\IEEEcompsocitemizethanks{
\IEEEcompsocthanksitem P. Habibi and A. Leon-Garcia are with The Edward S. Rogers Sr. Department of Electrical and Computer Engineering, University of Toronto, Toronto, ON M5S 3G4, Canada.
\IEEEcompsocthanksitem Corresponding author: P. Habibi (e-mail: pooyan.habibi@utoronto.ca).}}

\date{}

\begin{document}
\bstctlcite{IEEEtranBSTCTL:BSTcontrol}
\maketitle

\begin{abstract}
Maintaining service-level objectives (SLOs) on Kubernetes microservices
remains difficult because autoscalers observe coarse resource metrics,
recent SLO controllers often depend on custom telemetry, and
unconstrained agentic operators cannot safely mutate production
clusters. We present \sys{}, a guarded control plane for SLO-oriented
Kubernetes remediation. \sys{} builds an \otel-native causal resource
graph, assembles bounded \texttt{DiagnosisContext} objects, and exposes
a finite typed-action interface that separates planning from execution.
The same interface supports deterministic planners and an LLM-backed
planning harness, with deterministic schema checks, policy gates,
resource/disruption budgets, approval, and bounded execution forming
the safety substrate.

We evaluate \sys{} on a 4-node Kubernetes cluster using
DeathStarBench Social Network and Online Boutique. The evaluation
tests two forms of SLO-oriented control that resource autoscaling alone
does not provide: selecting the right remediation action and selecting
the right downstream target. For bad-image deployment regressions,
\sys{} selects \texttt{rollback\_canary} in every CPU-burn and
pure-latency run, five per flavor; HPA either scales the faulty image or
never triggers. For a downstream critical-path fault, the user-visible
breach appears at the frontend, but trace evidence identifies
\texttt{home-timeline-service} as the remediable bottleneck.
Deterministic \sys{} and a live approval-gated Sonnet harness target
that downstream service in every replicate, whereas HPA/resource-only
control never does. Additional experiments cover guarded placement
repair, Online Boutique portability, adversarial safety rejection,
offline multi-model replay, and KWOK-based control-plane scale
evidence. We release the controller, DSB-RegCtx corpus, workload
harnesses, safety tests, scale scripts, and figure artifacts:
\url{https://github.com/pooyan/arbiter}.
\end{abstract}

\begin{IEEEkeywords}
Kubernetes, microservices, service-level objectives, cloud resource
management, OpenTelemetry, distributed tracing, AIOps, safe agentic
systems.
\end{IEEEkeywords}

\section{Introduction}

Kubernetes is effective at maintaining declared infrastructure state,
but application SLOs are usually expressed outside that control loop. A
latency-sensitive request may traverse frontend gateways, application
services, databases, caches, service mesh sidecars, and external APIs,
while the cluster acts on lower-level objects such as pods,
deployments, resource requests, affinity rules, and replica counts. The
systems problem is therefore not only to detect an SLO violation, but
to translate user-visible evidence into the right bounded Kubernetes
action on the right target.

This translation is difficult for two reasons. First, operators often
overprovision services because they must reserve capacity for
worst-case demand, noisy neighbors, and poorly understood dependency
behavior. Second, SLO violations still occur despite overprovisioning
because tail latency can arise from qualitatively different causes:
critical-path shifts, placement and co-location interference, resource
pressure, or recent deployment changes. A high CPU signal may call for
resizing or scaling, a bad placement may call for eviction or topology
repair, and a bad image rollout may call for rollback rather than more
capacity. Single-signal policies such as CPU-based horizontal pod
autoscaling observe only one part of this decision.

Prior systems establish that SLO-aware resource control is a central
systems problem. FIRM showed the value of critical-path localization and
fine-grained reprovisioning~\cite{firm}; Autothrottle showed that
application-level SLO feedback can be decoupled from service-level
resource actuation~\cite{autothrottle}. The current cloud-native
environment, however, makes a different architecture possible: \otel{}
now provides a standard telemetry substrate for traces, metrics, logs,
profiles, and resource attributes~\cite{otelcollector}; Kubernetes
exposes richer resource and scheduling primitives, including the
scheduling framework, dynamic resource allocation, pod-level resource
declarations, autoscaling, and pressure stall information~\cite{k8ssched};
and agentic systems can synthesize heterogeneous evidence, runbooks, and
policy constraints, provided that their outputs are constrained by
verifiable execution mechanisms.

The design position of \sys{} is that SLO remediation should be treated
as a guarded translation problem. Telemetry and traces identify
candidate causes; planners propose finite typed actions; and only
deterministic validators and bounded controllers hold authority to
mutate the cluster. This separation turns agentic planning from an
unsafe autonomous operator into a replaceable hypothesis generator
behind a verifiable Kubernetes action interface. It also broadens SLO
control beyond autoscaling: some incidents require changing a resource
envelope, while others require moving a pod, repairing placement,
respecting disruption budgets, or rolling back a recent change.

This architecture rests on three principles: standardize the evidence
plane with \otel-compatible signals enriched by Kubernetes and node
metadata; separate reasoning from execution so planners propose actions
but never patch the cluster directly; and treat runtime control and
placement as one remediation surface, since the correct SLO action may
depend on both resource state and where a pod is running.

This paper makes four contributions:
\begin{enumerate}[leftmargin=*,label=(\arabic*)]
  \item \textbf{An \otel-native \texttt{DiagnosisContext} graph
  substrate} that correlates \otel{} telemetry with Kubernetes state,
  placement metadata, node pressure, and recent-change history to
  produce a bounded cross-layer evidence object for planners
  (Fig.~\ref{fig:graph-plane},
  \S\ref{sec:graph-plane}).
  \item \textbf{A typed Kubernetes remediation interface}: structured
  \texttt{DiagnosisContext} JSON in, finite-vocabulary typed action JSON
  out. We evaluate deterministic and single-call agentic planners on the
  same interface (\S\ref{sec:agents}, \S\ref{sec:agentic-ablation}).
  \item \textbf{A deterministic safety validator and bounded executor
  for typed remediation plans.} The safety layer enforces schema,
  allowed action types, deny-list policy, resource/disruption budgets,
  approval gating, concurrent-intervention caps, ResourceQuota checks,
  PodDisruptionBudget feasibility for evictions, and optional
  maintenance windows outside the LLM. This separation grounds the
  paper's calibrated thesis: \emph{the deterministic substrate is what
  makes agentic planning safe; agentic planning is what makes the
  deterministic substrate extensible.}
  \item \textbf{A reproducible live-workload and scale evaluation
  corpus}: DSB-RegCtx captured contexts and replay scripts, DSB Social
  Network and Online Boutique injected-fault harnesses, and separate
  KWOK plus kube-burner/ClusterLoader2 scale-and-churn workflows.
  Runtime results come from the live 4-node cluster; fake-node scale and
  real-cluster churn are reported as a distinct control-plane evidence
  track rather than as microservice-runtime claims.
\end{enumerate}

The rest of the paper is organized as follows. Section~\ref{sec:background}
motivates the Kubernetes SLO-control gap; Section~\ref{sec:problem}
defines the control problem; Sections~\ref{sec:overview}--\ref{sec:safety}
present the \sys{} architecture, diagnosis interface, and safety model;
Sections~\ref{sec:implementation}--\ref{sec:results} describe the
implementation and evaluation; and Sections~\ref{sec:related}--\ref{sec:conclusion}
discuss related work, claim boundaries, and conclusions.

\section{Background and Motivation}
\label{sec:background}

\subsection{The SLO Control Gap in Kubernetes}

Kubernetes gives operators a portable control surface for placement,
replication, and resource management. Pods declare resource requests and
constraints, the scheduler maps feasible pods to nodes, and controllers
drive the observed cluster toward declared state~\cite{k8ssched}. These
interfaces are intentionally object-centric: they reason about pods,
nodes, deployments, quotas, and policies. Application SLOs, however, are
defined over user-visible request behavior. The result is an evidence
mismatch. The cluster exposes useful knobs, but the native control loop
does not directly know a request's critical path, recent deployment
history, co-location context, or end-to-end tail latency.

For microservices, the same p99 latency breach may require different
and sometimes mutually exclusive actions. Adding replicas can help a
throughput-limited service, but may amplify interference if replicas are
placed on overloaded nodes. Raising CPU limits can help a throttled
container, but not a downstream queue, network path, database lock, or
bad image rollout. Evicting a pod can repair co-location interference,
but may be harmful when locality or warm cache state dominates. A useful
SLO controller therefore needs to translate combined application and
cluster evidence into the right bounded action on the right target.

\subsection{Lessons from Prior SLO Controllers}

Prior SLO controllers are useful here not as direct templates, but as
design constraints for Kubernetes remediation. FIRM implies that
localization should follow the request critical path rather than the
service with the largest isolated latency~\cite{firm}. High variance
and resource contention along the critical path are more actionable than
a flat ranking of slow services. Autothrottle implies that practical
resource control benefits from decoupling application-level SLO feedback
from service-level actuation~\cite{autothrottle}; a global SLO signal
can guide local controllers without turning the whole application into
one monolithic control problem.

\sys{} takes these lessons into a different substrate. In Kubernetes,
the remediation surface is not only CPU allocation. It includes
replicas, resource envelopes, priority and throttling, placement repair,
rollbacks, topology constraints, and dynamic resource allocation.
Consequently, the central problem is not choosing between deterministic
control and agentic reasoning. It is building a typed translation layer
that can carry critical-path and cluster evidence to a finite action
interface, while keeping permission to mutate the cluster inside
deterministic validators and bounded controllers.

\subsection{Why \otel{} Changes the Design Space}

Earlier microservice SLO systems often required custom tracing, custom
telemetry schemas, or benchmark-specific instrumentation. \otel{} changes
this tradeoff by providing a vendor-neutral substrate for traces,
metrics, logs, profiles, resource attributes, semantic conventions, and
collector pipelines. In Kubernetes, \otel{} Collector components can
attach namespace, pod, deployment, node, and other resource metadata to
telemetry records~\cite{otelcollector}. This identity enrichment is what
turns observability data into controller evidence: a root-span latency
sample can be correlated with workload ownership, image versions, node
pressure, placement, and recent Kubernetes changes.

This portability matters for \sys{} because the controller should not
own application instrumentation or depend on one monitoring stack.
\sys{} ingests OTLP trace data; resource and node metrics are read from
the Kubernetes metrics API. Metrics, logs, and profile signals are
carried by the same collector pipeline design but are not consumed by
the current controller. Its graph then assembles a
bounded \texttt{DiagnosisContext}: a structured cross-layer evidence
object that planners can inspect and validators can audit, instead of a
private collection of benchmark-specific signals.

\section{Problem Statement}
\label{sec:problem}

We consider a Kubernetes cluster running one or more latency-sensitive microservice applications. Each application defines one or more SLOs over request classes, such as p99 latency below a target threshold over a rolling window, error rate below a threshold, or availability above a threshold (the evaluated implementation samples p99 latency). The cluster contains heterogeneous nodes and may include accelerators, network-attached storage, service mesh components, sidecars, and background workloads.

At each control interval, the system observes telemetry streams and cluster state. The design space of actions includes: adjust replica counts; adjust container or pod resource requests and limits; modify throttling or priority of lower-priority services; move or evict pods through descheduling; influence future placements through scheduler scoring or constraints; allocate or release dynamic resources; change traffic splitting for canaries or rollbacks; or take no action and continue observing; the evaluated controller implements the four-action subset of Table~\ref{tab:actions}. In Kubernetes controller terms, \emph{current state} is the bounded \texttt{DiagnosisContext} plus live gate inputs; \emph{desired state} is one typed, bounded transition on one target; and \texttt{preState} is its rollback snapshot.

The objective is to minimize SLO violation frequency and duration while maximizing resource efficiency and avoiding unsafe interventions. \sys{} must operate under five constraints:
\begin{itemize}[leftmargin=*]
  \item \textbf{Safety:} actions must respect quotas, priorities, disruption budgets, tenant boundaries, and administrator policies.
  \item \textbf{Stability:} actions must avoid oscillation and cascading remediation loops.
  \item \textbf{Auditability:} each action must be explainable in terms of observed evidence, expected effect, and rollback condition.
  \item \textbf{Timeliness:} diagnosis and action selection must complete within the time scale of manageable SLO incidents.
  \item \textbf{Portability:} telemetry and control interfaces should be compatible with standard Kubernetes and \otel{} deployments when possible.
\end{itemize}

Non-goals are equally important. \sys{} does not attempt to replace Kubernetes scheduling, admission control, or observability systems. It does not assume an LLM can learn resource control from raw telemetry. It does not directly solve application bugs such as memory leaks or correctness faults, although it can identify evidence suggesting that resource remediation is unlikely to help.

\section{\sys{} Control Architecture}
\label{sec:overview}

\sys{} is organized around one invariant: reasoning is replaceable, but
mutation authority is not. A planner may be deterministic, agentic, or a
future hybrid, but it can only emit a finite typed action. The same
validator, contextual safety gates, bounded executor, pre-state capture,
and rollback machinery decide whether that action reaches the
Kubernetes API.

\begin{figure*}[t]
\centering
\includegraphics[width=\textwidth]{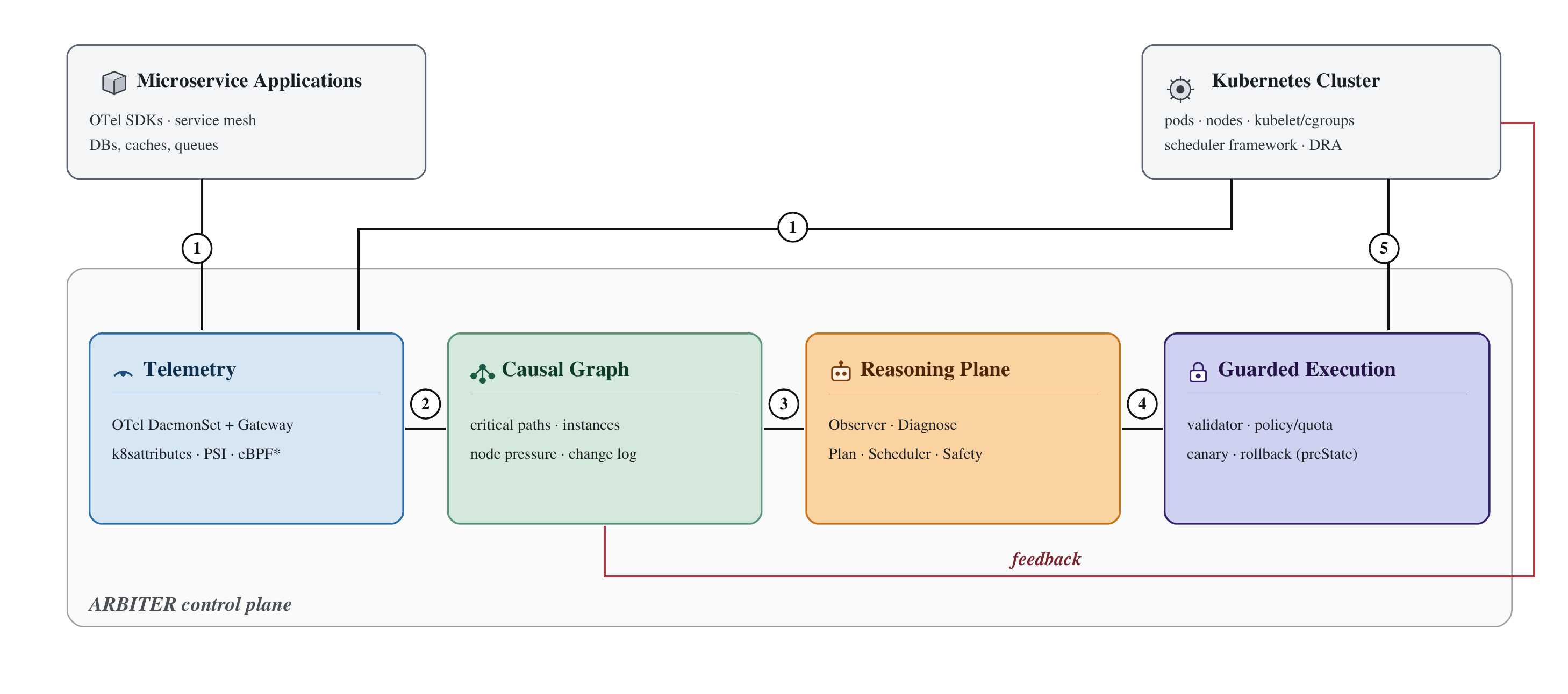}
\caption{\sys{} architecture. Telemetry from applications and the cluster
(\protect\circled{1}) is normalized into a causal resource graph
(\protect\circled{2}). The reasoning plane reads compact graph contexts and
emits typed JSON plans (\protect\circled{3}). The execution plane validates each
plan and patches the Kubernetes API only through typed controllers
(\protect\circled{4}--\protect\circled{5}). Outcome metrics flow back into the
graph (red, dashed). The graph and execution planes are detailed in
Figs.~\ref{fig:graph-plane} and~\ref{fig:exec-plane};
\S\ref{sec:tel-plane}--\S\ref{sec:safety} describe all four planes.}
\label{fig:architecture}
\end{figure*}

The control plane has four planes: telemetry, causal graph, reasoning,
and guarded execution. Figure~\ref{fig:architecture} sketches the loop
end-to-end; the graph and execution planes are detailed in
Figs.~\ref{fig:graph-plane} and~\ref{fig:exec-plane}. Telemetry is collected near the workload,
enriched with Kubernetes identity, and ingested by the controller. The
graph correlates request critical paths with placement, node pressure,
and recent-change history, then assembles bounded
\texttt{DiagnosisContext} JSON. The reasoning plane reads that context
and emits a typed plan; the execution plane validates the plan, captures
pre-state when needed, patches only the declared target, and rolls back
if the action's rollback trigger fires inside the observation window.
Outcome records are annotated back into the graph as intervention
vertices when a plan reaches a terminal phase, so operators and future
planners can inspect the effect of prior interventions; the current
deterministic planner does not consume them.

\subsection{Telemetry Plane}
\label{sec:tel-plane}

The telemetry plane collects and normalizes signals from the application and cluster. It uses \otel{} as the primary transport and schema layer. Application services can emit traces, metrics, logs, and profiles through standard SDKs, auto-instrumentation, service mesh integration, or sidecars, but the evaluated controller ingests only OTLP trace data and reads resource and node metrics from the Kubernetes metrics API. Kubernetes and node signals can also be collected through \otel{} Collector receivers, Prometheus scraping, Kubernetes object watchers, kubelet statistics, PSI metrics, and optional eBPF/perf receivers (syscall latency, TCP retransmits, run-queue delay, block I/O latency, cgroup-level resource stalls); this instrumentation surface is reserved in the context schema and not consumed by the evaluated controller. \sys{} is designed to operate in a standard \otel/Kubernetes environment and to gain diagnosis confidence when deeper node telemetry is available.

Telemetry records are enriched with Kubernetes resource-identity
attributes such as namespace, deployment, replica set, pod UID,
container name, node name, image version, workload owner, and tenant
label. This semantic enrichment is the basis for cross-layer
correlation: request-level latency and error observations can be
associated with the exact workload, image, pod, node, and tenancy
context that produced them. The deployed shape is a per-node DaemonSet
collector that owns receivers close to the workload, a cluster-wide
gateway that performs metadata enrichment (\texttt{k8sattributes}) and
tail sampling, and a controller-side OTLP/gRPC ingest endpoint that
feeds the causal graph; collector configurations are reproducible
artifacts in the open-source repository.

At each control interval the controller samples the SLO window,
recent-change buffer, critical-path lookback, and live pod/node
metrics; sampler counters record dropped and matched request classes.

\subsection{Causal Resource Graph}
\label{sec:graph-plane}

The graph plane is the evidence model that turns normalized telemetry
into planner-readable intervention state. Conceptually, it is a typed
intermediate representation for cloud control: not a monitoring
warehouse, but a compact structure in which request behavior, workload
identity, placement, resource pressure, and recent changes become
co-referent evidence. It maintains a time-indexed heterogeneous graph
whose vertices carry stable operational identity rather than only metric
labels. Request and span vertices preserve trace provenance; workload
vertices bind pods to owners, namespaces, and nodes; node vertices carry
placement identity, while pressure/capacity data is sampled from the
metrics API per control interval rather than stored on the vertex;
richer resource-domain vertices (I/O, network paths,
devices, storage) are schema extensions. This identity layer is the
mechanism that converts an SLO symptom into an actionable control
surface: a root-span latency sample can be resolved to the downstream
span that dominates the request, the pod that executed it, the
Deployment that owns the pod, and the policy context that constrains any
repair.

We use ``causal'' in an operational sense. The graph does not claim
formal causal identification from observational data alone. Instead, it
preserves intervention-relevant provenance: which request path was
affected, which runtime object produced the signal, which placement or
resource condition was contemporaneous, and which recent change altered
the workload. The graph therefore separates \emph{structural} edges
(trace parent-child relations, ownership, and scheduling) from
\emph{evidential} edges. The implemented evidential
edge is pod co-location, annotated with the co-residency window,
overlap measurement, temporal support, and confidence;
latency-delta, pressure-correlation, and recent-change-proximity edges
are extensions of the same edge class.

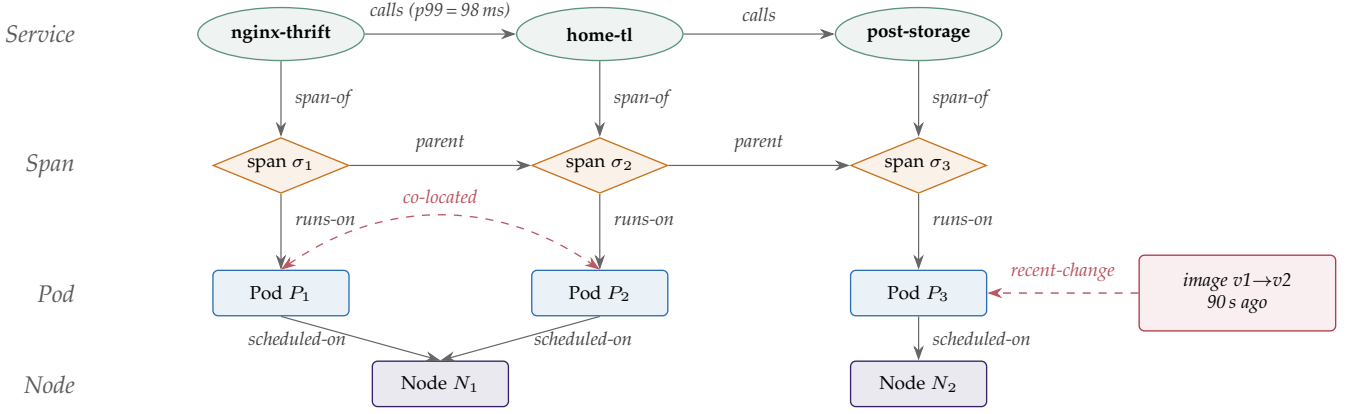
\begin{figure*}[t]
\centering
\begin{tikzpicture}[
  font=\scriptsize,
  >={Stealth[length=2mm,width=1.4mm]},
  layerlabel/.style={font=\small\itshape, text=black!65, anchor=east},
  svc/.style={ellipse, draw=graphcolor, fill=graphcolor!10, line width=0.5pt,
    minimum width=22mm, minimum height=7mm, inner sep=1pt, font=\scriptsize\bfseries},
  pod/.style={rectangle, rounded corners=2pt, draw=telcolor, fill=telcolor!10,
    line width=0.5pt, minimum width=18mm, minimum height=6mm, inner sep=1pt},
  node1/.style={rectangle, rounded corners=2pt, draw=execcolor, fill=execcolor!10,
    line width=0.5pt, minimum width=18mm, minimum height=6mm, inner sep=1pt},
  span/.style={diamond, aspect=2, draw=agentcolor, fill=agentcolor!10,
    line width=0.5pt, inner sep=1pt, minimum width=18mm, minimum height=6mm},
  chg/.style={rectangle, rounded corners=2pt, draw=feedbackcolor, fill=feedbackcolor!8,
    line width=0.5pt, minimum width=26mm, minimum height=10mm, inner sep=2pt,
    align=center, font=\scriptsize\itshape},
  e/.style={->, draw=black!60, line width=0.45pt},
  ev/.style={->, draw=feedbackcolor!85, line width=0.55pt, dashed},
  elabel/.style={font=\scriptsize\itshape, text=black!75},
  evlabel/.style={font=\scriptsize\itshape, text=feedbackcolor!90},
]
  % --- Service row ---
  \node[svc] (s1) at (0,0)              {nginx-thrift};
  \node[svc, right=20mm of s1] (s2)     {home-tl};
  \node[svc, right=20mm of s2] (s3)     {post-storage};

  % --- Span row ---
  \node[span, below=10mm of s1] (sp1)   {span $\sigma_1$};
  \node[span, below=10mm of s2] (sp2)   {span $\sigma_2$};
  \node[span, below=10mm of s3] (sp3)   {span $\sigma_3$};

  % --- Pod row ---
  \node[pod, below=10mm of sp1] (p1)    {Pod $P_1$};
  \node[pod, below=10mm of sp2] (p2)    {Pod $P_2$};
  \node[pod, below=10mm of sp3] (p3)    {Pod $P_3$};

  % --- Recent-change callout, same row as pods, right of P3 ---
  \node[chg, right=20mm of p3] (ch) {image v1$\to$v2\\\scriptsize 90\,s ago};

  % --- Node row ---
  \node[node1] (n1) at ($(p1)!0.5!(p2)+(0,-12mm)$) {Node $N_1$};
  \node[node1] (n2) at ($(p3)+(0,-12mm)$)          {Node $N_2$};

  % --- Layer labels (faint italic on the left) ---
  \coordinate (lcol) at ($(s1.west)+(-15mm,0)$);
  \node[layerlabel] at (lcol |- s1)  {Service};
  \node[layerlabel] at (lcol |- sp1) {Span};
  \node[layerlabel] at (lcol |- p1)  {Pod};
  \node[layerlabel] at (lcol |- n1)  {Node};

  % --- Service-level edges (calls) ---
  \draw[e] (s1) -- (s2) node[elabel, midway, above=1.5pt] {calls (p99\,=\,98\,ms)};
  \draw[e] (s2) -- (s3) node[elabel, midway, above=1.5pt] {calls};

  % --- Span-of edges (Service -> Span vertical) ---
  \draw[e] (s1) -- (sp1) node[elabel, pos=0.5, anchor=west, xshift=2pt] {span-of};
  \draw[e] (s2) -- (sp2) node[elabel, pos=0.5, anchor=west, xshift=2pt] {span-of};
  \draw[e] (s3) -- (sp3) node[elabel, pos=0.5, anchor=west, xshift=2pt] {span-of};

  % --- Parent chain (Span -> Span horizontal) ---
  \draw[e] (sp1.east) -- (sp2.west) node[elabel, midway, above=1.5pt] {parent};
  \draw[e] (sp2.east) -- (sp3.west) node[elabel, midway, above=1.5pt] {parent};

  % --- Runs-on edges (Span -> Pod vertical). Label sits at pos=0.35
  % (closer to the span row) so it does not collide with the co-located
  % red curve that emerges from p1.north / p2.north.
  \draw[e] (sp1) -- (p1) node[elabel, pos=0.35, anchor=west, xshift=2pt] {runs-on};
  \draw[e] (sp2) -- (p2) node[elabel, pos=0.35, anchor=west, xshift=2pt] {runs-on};
  \draw[e] (sp3) -- (p3) node[elabel, pos=0.35, anchor=west, xshift=2pt] {runs-on};

  % --- Scheduled-on edges (Pod -> Node) ---
  \draw[e] (p1.south) -- (n1.north) node[elabel, pos=0.5, anchor=east, xshift=-2pt] {scheduled-on};
  \draw[e] (p2.south) -- (n1.north) node[elabel, pos=0.5, anchor=west, xshift=2pt] {scheduled-on};
  \draw[e] (p3.south) -- (n2.north) node[elabel, pos=0.5, anchor=west, xshift=2pt] {scheduled-on};

  % --- Co-located edge: bends ABOVE the pod row, out of the pod->node arrow zone ---
  \draw[<->, draw=feedbackcolor!85, line width=0.5pt, dashed]
       (p1.north) to[bend left=35]
       node[evlabel, above=1.5pt] {co-located}
       (p2.north);

  % --- Recent-change edge: horizontal arrow into P3.east ---
  \draw[ev] (ch.west) -- (p3.east)
       node[evlabel, midway, above=1pt] {recent-change};
\end{tikzpicture}
\caption{Causal resource graph (\protect\circled{2} of
Fig.~\ref{fig:architecture}), schema and one populated example. The four
layers (Service, Span, Pod, Node) are correlated through structural edges
(\texttt{span-of}, \texttt{parent}, \texttt{runs-on}, \texttt{scheduled-on}); a
co-location edge ($P_1, P_2$ on $N_1$) and a recent-change edge (image flip
90\,s ago on $P_3$'s Deployment, dashed red) are the evidential edges that
distinguish a placement incident from a deployment regression.}
\label{fig:graph-plane}
\end{figure*}

Figure~\ref{fig:graph-plane} shows the schema in miniature. The same
cluster entity can appear simultaneously as workload, placement, and
runtime context: a span on a request critical path resolves to a specific
pod, the pod resolves to a Deployment and node, and the node anchors
placement identity. Co-location is expressed as evidential edges between
the pods it hosts. In the implementation reported here,
the graph produces critical-path, placement and co-location, resource,
and recent-change evidence; the deterministic planner consults a
deliberately narrow subset of these fields, and remaining schema fields
are reserved for future signals.

Given this representation, \sys{} asks intervention-oriented graph
queries rather than raw monitoring queries. The graph identifies request
critical paths that dominate an SLO breach, attributes tail latency to
the dominant critical-path Deployment, and checks whether recent
changes fall within a fixed proximity window (240\,s) of the symptom.
The controller records intervention outcomes into the graph as
vertices, and separately suppresses duplicate in-flight plans per
target and action type via labeled \texttt{RemediationPlan} queries. The
implementation exposes these queries through \texttt{DiagnosisContext};
the deterministic planner includes a critical-path branch that can
target a dominant downstream Deployment rather than the breached
frontend when the graph evidence points there.

\begin{figure*}[t]
\centering
\includegraphics[width=\textwidth]{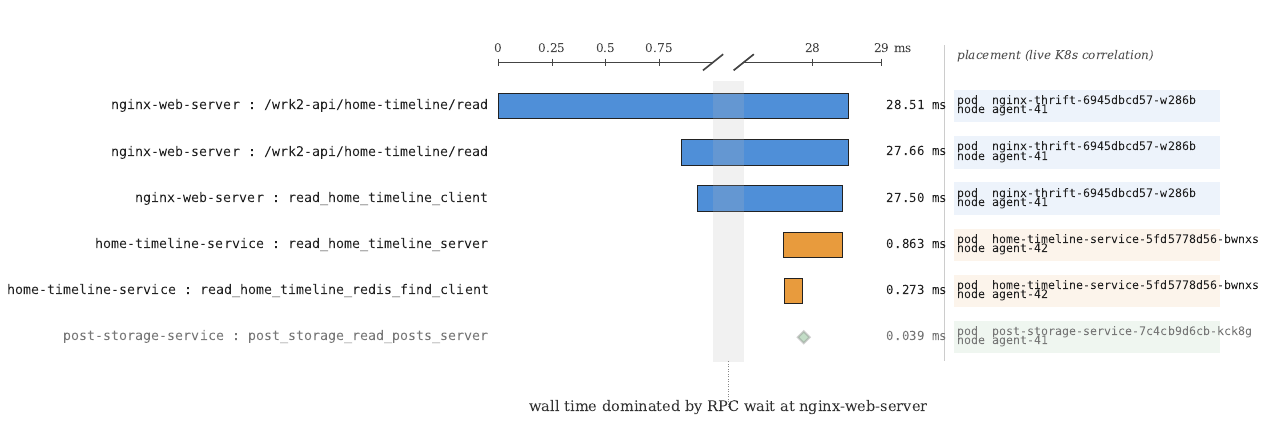}
\caption{Critical-path query output for one DSB
\texttt{read-home-timeline} request from a real OTLP trace. Bars show
per-span wall time with a diagonal scale break; indentation marks
parent--child depth; the right column shows live Kubernetes placement
correlated with the trace by the controller.}
\label{fig:causal-graph}
\end{figure*}

Figure~\ref{fig:causal-graph} shows one such query output for a real DSB
\texttt{read-home-timeline} request, recorded from \sys's live OTLP ingest
path against the DeathStarBench Social Network benchmark. The chain
crosses three services and is six spans deep, but the important
observation is not depth alone: wall time is concentrated where
\texttt{nginx-web-server} waits on a downstream RPC, while the leaf
services' own work contributes microseconds. The graph materializes this
path evidence as reusable diagnosis context alongside placement and
recent-change history (Listing~\ref{lst:context}). The critical-path
branch then uses the Deployment identity attached to each path step to
select a remediable backend when it dominates a breached request's tail
latency; resource-only HPA cannot make that distinction because it sees
neither the trace path nor the downstream target.\footnote{Source data for
Figs.~\ref{fig:graph-plane}--\ref{fig:causal-graph} and
Listing~\ref{lst:context} were captured from a live \sys{} deployment on
the 4-node testbed cluster; the trace, the live Kubernetes placement at
trace time, and the diagnosis context are reproduced verbatim from the
captured artifacts. All figure-rendering scripts and captured artifacts
are available with the open-source release.\label{fn:artifacts}}

For each request class, \sys{} maintains a bounded streaming summary
rather than retaining unbounded trace history. Finished OTLP spans are
folded into the current control window; the extractor accumulates
per-service exclusive time along the longest child chain from the root
and records the dominant service leg in \texttt{critical\_path[]}. From
the same window, the graph derives co-location evidence as annotated
evidential edges; further window-derived signals (queueing delay, retry
amplification, instance skew, downstream blocking probability) are
reserved extensions of the same annotation schema. The implemented signals
are evidence features for ranking hypotheses and candidate interventions.
They do not grant permission to mutate the cluster; mutation authority remains with
the typed validator and executor. Table~\ref{tab:graph-cost} reports the
graph plane's measured cost at the deployed bound (20\,000 spans, 600\,s
window; synthetic workload with 4 nodes, 40 pods, 12 services) on a
development machine. Ingest-path operations are sub-2\,\textmu s
(core upsert) to $\approx$20\,\textmu s per span including structural and
co-location edge maintenance; control-loop queries run per breach
evaluation rather than per span, and their linear span scans are
inherited unchanged from the store the graph replaces.

\begin{table}[t]
\centering
\small
\caption{Causal-graph cost at the deployed bound (20\,000 spans, 600\,s
window), Go microbenchmarks on a development machine (Xeon Gold 5118).}
\label{tab:graph-cost}
\begin{tabularx}{\columnwidth}{@{}X r r@{}}
\toprule
Operation & Latency & Alloc/op \\
\midrule
Core vertex upsert (at capacity) & 1.4\,\textmu s & 155\,B \\
Span ingest incl.\ edge upserts & 19.6\,\textmu s & 3.8\,KB \\
Downstream closure (depth $\leq 6$) & 7.2\,\textmu s & 1.8\,KB \\
Critical-path extraction & 7.7\,ms & 4.8\,MB \\
Slowest-root query (60\,s lookback) & 19.8\,ms & 10.7\,MB \\
Hot-instance ranking & 7.8\,ms & 4.8\,MB \\
\midrule
Resident heap at capacity & \multicolumn{2}{r}{70.2\,MiB
(20\,056 vertices, 56\,886 edges)} \\
\bottomrule
\end{tabularx}
\end{table}

\subsection{Reasoning Plane}
\label{sec:agents}

The reasoning plane is the semantic boundary where graph evidence is
compiled into Kubernetes intent. It is deliberately narrower than an
autonomous operator and richer than a scalar autoscaler: it does not
inspect raw telemetry streams or patch the API server, but it does
interpret a bounded \texttt{DiagnosisContext} as an intervention problem.
The output is exactly one finite-vocabulary JSON action whose fields
name the target, action type, preconditions, expected effect,
observation window, rollback trigger, confidence, and safety budget.
This makes planning a replaceable translation step. Deterministic
rules, single-call LLMs, or future multi-agent planners can occupy the
same reasoning slot, while validation and execution remain unchanged.

This boundary is also what makes the evaluation comparable. In the
primary live measurements, the deterministic planner consults
\texttt{recent\_changes}, pod-CPU, node-pressure, and
\texttt{critical\_path} fields. In the offline ablation and the live
approval-gated Sonnet harness, the model consumes the same
\texttt{DiagnosisContext} and emits the same typed action shape
(\S\ref{sec:prelim-regression}, \S\ref{sec:agentic-ablation}). The
comparison is therefore about action selection over identical evidence,
not about granting different planners different observability,
privilege, or actuation channels.

\paragraph*{Deterministic planner} The deterministic planner is an
executable specification of the remediation predicates used in the live
experiments. Each branch is intentionally narrow and evidence-gated. A
deployment-regression predicate detects an image flip in
\texttt{recent\_changes} on the breached Deployment within a 240\,s
window and emits \texttt{rollback\_canary}. A critical-path predicate
redirects the target to the dominant downstream Deployment when
\texttt{critical\_path} disagrees with the breached frontend. When no
stronger predicate applies, the planner considers bounded resource and
placement actions: \texttt{scale\_out} (Replicas$=$1),
\texttt{resize\_cpu} (CPU$=$0.1), and \texttt{deschedule\_one}
(Evictions$=$1, only when NodeCPU$\geq$70\%). Among candidates the
scorer picks $\arg\max_{a}\,\textsc{conf}(a) -
0.2\!\cdot\!\textsc{cost}(a)$, where the cost term is the aggregate
safety budget; this is the arithmetic referenced by the long-tail replay
analysis of \S\ref{sec:agentic-ablation}. The point of these branches is
not to claim an optimal scheduler. They are a controlled test of whether
the graph context can disambiguate rollback, critical-path targeting,
resource adjustment, and placement repair cases that resource-only
baselines collapse into the same response. \S\ref{sec:prelim-regression}
shows the regression branch firing on a real cluster;
\S\ref{sec:critical-path-results} evaluates the critical-path branch
under a live downstream fault.

\paragraph*{Single-call agentic planner} The evaluated agentic form is
minimal by design: one prompt, one bounded \texttt{DiagnosisContext}, and
one typed-action JSON object. The model is treated as an untrusted
semantic planner, not as a Kubernetes principal. Its task is to map
structured evidence to a schema-conforming intervention hypothesis; the
validator later decides whether that hypothesis is admissible. The
offline replay in \S\ref{sec:agentic-ablation} evaluates three pinned
models --- Sonnet 4.6 (\texttt{claude-sonnet-4-6}), Haiku 4.5
(\texttt{claude-haiku-4-5-20251001}), and Opus 4.7
(\texttt{claude-opus-4-7}) --- against captured contexts the
deterministic planner already saw plus one synthesized noisy-neighbor
scenario the deterministic rule set does not cover. The live-gated
Sonnet harness uses the same interface, but wraps the model output as an
agent-labeled \texttt{RemediationPlan} that still requires deterministic
validation and operator approval before execution. Richer multi-role
planning can be substituted into this slot without changing the safety
or execution planes, but it is outside the evaluated system.

\subsection{Guarded Execution Plane}
\label{sec:exec-plane}

The execution plane is the mutation boundary of \sys{}. It is
deliberately non-agentic: planners may propose intent, but only this
plane holds the authority to translate intent into Kubernetes API
patches. The central abstraction is a typed intervention envelope. Each
action names a \texttt{target}, an \texttt{action\_type}, explicit
\texttt{preconditions}, an \texttt{expected\_effect}, a
\texttt{confidence}, a \texttt{rollout\_scope}, a bounded
\texttt{safety\_budget}, an \texttt{observation\_window}, and a
\texttt{rollback\_trigger}. This envelope makes every mutation
auditable before execution, monitorable after execution, and comparable
across deterministic and agentic planners.

\begin{figure}[!t]
\centering
\includegraphics[width=\columnwidth]{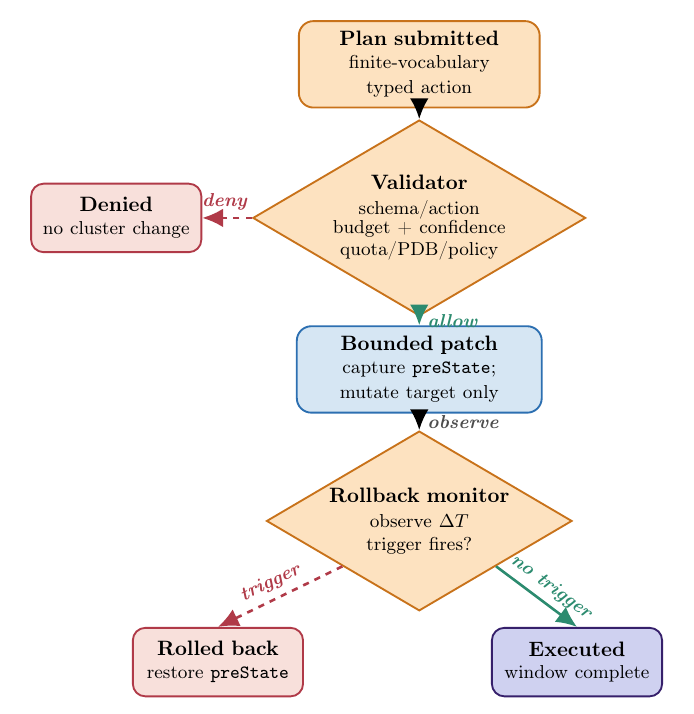}
\caption{Guarded execution state machine
(\protect\circled{4}--\protect\circled{5} of Fig.~\ref{fig:architecture}).
The control path is validate $\rightarrow$ gate $\rightarrow$ patch
$\rightarrow$ observe $\rightarrow$ retain/rollback. Every
\texttt{RemediationPlan} traverses the same deterministic mutation envelope:
schema/action validation, budget and confidence checks, contextual gates
(quota, PDB, deny-list, maintenance-window, concurrent-intervention),
optional \texttt{preState} capture, bounded patching, and rollback
monitoring. Traffic-level canary scope is outside the evaluated action set
(see \S\ref{sec:safety}).}
\label{fig:exec-plane}
\end{figure}

Table~\ref{tab:actions} shows the four actions implemented and evaluated
here. The set is intentionally finite: the paper evaluates the safety
substrate under bounded replica, resource, eviction, and rollback
actions, while leaving richer scheduling, traffic, and interference
actions to reuse the same interface. Figure~\ref{fig:exec-plane} shows
the implemented state machine. A plan is admitted by the safety
validator, pre-state is captured for reversible mutations, the bounded
patch is applied only to the declared target, and the controller restores
the captured state if the rollback trigger fires inside the observation
window. \S\ref{sec:safety} specifies the deterministic gates that stand
between planner reasoning and cluster mutation.

\begin{table}[t]
\centering
\small
\caption{Implemented and evaluated typed action vocabulary.}
\label{tab:actions}
\begin{tabularx}{\columnwidth}{@{}p{0.34\columnwidth} p{0.18\columnwidth} X@{}}
\toprule
Action & Budget cap & Kubernetes interface \\
\midrule
\texttt{scale\_out} & replicas $\leq 1$ & Deployment scale subresource \\
\addlinespace[3pt]
\texttt{resize\_cpu} & CPU $\leq 0.1$ & Deployment pod-template resource patch \\
\addlinespace[3pt]
\texttt{deschedule\_one} & evictions $\leq 1$ & Pod eviction / descheduler-style repair \\
\addlinespace[3pt]
\texttt{rollback\_canary} & no capacity increase & Deployment image rollback from \texttt{RecentChanges} \\
\bottomrule
\end{tabularx}
\end{table}

Execution proceeds as a guarded commit sequence. First, the validator
checks the proposal against static policy: allowed action type,
deny-listed targets, safety budget, and confidence. Second, just before
execution, contextual gates bind the decision to current cluster state:
namespace ResourceQuota headroom for capacity-changing actions,
PodDisruptionBudget feasibility before \texttt{deschedule\_one}, the
configured maintenance window, and the number of concurrently executing
\texttt{RemediationPlan}s. Third, the controller persists a pre-state
snapshot for reversible mutations, including the current replica count
and container resource requests. Fourth, it applies the bounded patch via
the Kubernetes API to the declared target only. Fifth, for actions with
implemented rollback monitors (\texttt{scale\_out},
\texttt{resize\_cpu}), it observes pod readiness during the window and
restores pre-state if readiness deteriorates beyond the configured grace
period. Traffic-level canary scope and aggregate-metric verdict gates
are left to later action modules behind the same interface.

\subsection{Guarded Control Loop}

\sys's control loop is event-driven and rate-limited, triggered by SLO
risk, active SLO violation, topology/resource change, or completion of
an earlier bounded action. Algorithm~\ref{alg:control} presents the loop
as an authorization pipeline: aligned evidence and graph state produce
candidate interventions, but only validated actions enter execution.

\begin{algorithm}[H]
\small
\caption{Guarded SLO control loop.}
\label{alg:control}
\begin{algorithmic}[1]
\Require SLO set $\mathcal{S}$, telemetry windows $\mathcal{W}$, Kubernetes
state $K$, policy $\Pi$, action history $\mathcal{H}$
\Ensure typed actions submitted only after validation, policy gates, and
rollback guards
\State $W \gets \Call{AlignSignals}{\mathcal{W}}$
  \Comment{OTLP traces, resource metrics, change events}
\State $Q \gets \Call{ScoreQuality}{W}$
\State $G \gets \Call{UpdateGraph}{}$
  \Comment{updated asynchronously from OTLP span ingest}
\State $\Sigma \gets \Call{DetectSymptoms}{\mathcal{S}, W}$
  \Comment{active or imminent SLO breach}
\ForAll{$\sigma \in \Sigma$}
  \State $C \gets \Call{ExtractContext}{G, \sigma}$
  \State $H \gets \Call{RankHypotheses}{C, Q}$
  \State $P \gets \Call{ProposeActions}{C, H}$
  \State $P \gets P \cup \Call{AddPlacementCandidates}{C, P}$
  \State $P \gets P \cup \Call{AddRuntimeCandidates}{C, P}$
  \State $a^\star \gets \Call{Score}{P, \Pi, \mathcal{H}}$
    \Comment{deterministic scorer}
  \State $a^\star \gets \Call{Validate}{a^\star, \Pi, K}$
  \If{$a^\star.\textit{decision} = \mathsf{deny}$}
     \State \Call{Record}{$a^\star$}; \textbf{continue}
  \ElsIf{$a^\star.\textit{decision} = \mathsf{require\_approval}$}
     \State \Call{Emit}{$a^\star$}; \textbf{continue}
  \EndIf
  \State $o \gets \Call{ExecuteBounded}{a^\star}$
    \Comment{captures \texttt{preState}}
  \If{$a^\star.\textit{type} \in \{\mathsf{scale\_out}, \mathsf{resize\_cpu}\}$}
     \State \Call{MonitorReadiness}{$a^\star, o$}
       \Comment{within $[\textit{grace}, \textit{window}]$}
     \If{\textit{degraded}}
        \State \Call{Rollback}{$a^\star, o$}
          \Comment{restores \texttt{preState}}
     \Else
        \State \Call{RecordSuccess}{$a^\star, o$}
     \EndIf
  \Else
     \State \Call{RecordSuccess}{$a^\star, o$}
       \Comment{no rollback monitor implemented for this action type}
  \EndIf
  \State $\mathcal{H} \gets \mathcal{H} \cup \{(a^\star, o)\}$;
         \Call{Annotate}{$G, a^\star, o$}
\EndFor
\end{algorithmic}
\end{algorithm}

The algorithm makes the research boundary explicit. Lines~1--4 build
the evidence frame and symptom set; lines~6--10 extract context and
propose runtime plus placement candidates; lines~11--17 perform scoring,
validation, denial, and approval gating. Only line~18 reaches the
bounded executor; line~29 records outcomes back into history and graph
annotations. Three guarantees follow. \textbf{Action validity}: every
patch is type-checked by \texttt{Action.Validate()} and the
\texttt{SafetyValidator} before execution. \textbf{Bounded blast radius}:
the four-axis safety budget is enforced regardless of planner output, so
out-of-budget requests reach \texttt{deny} before any Kubernetes patch.
\textbf{Reversibility}: \texttt{preState} is captured at execution time
and is the rollback target on line~22 for \texttt{scale\_out} and
\texttt{resize\_cpu}; actions without rollback monitors are
additionally bounded by the validator and, like all evaluated actions,
approval-gated (\S\ref{sec:safety}).

\section{Safety Model}
\label{sec:safety}

\sys{} treats every planner output as an untrusted advisory proposal
until it crosses a deterministic admissibility boundary. Deterministic
rules and agentic planners share this boundary: neither holds Kubernetes
mutation authority nor can issue arbitrary API patches. Both emit typed
actions from the finite schema in Table~\ref{tab:actions}, each carrying
preconditions, a resource budget, an observation window, and an explicit
rollback trigger.

The safety model has three layers. First, \emph{schema and policy
admissibility}: before execution, each action passes through the
\texttt{SafetyValidator}, which enforces typed-schema validity
(\texttt{Action.Validate()}), an allowed-action allow-list, an operator
deny-list over targets, a four-axis safety budget
(\texttt{MaxReplicasDelta}, \texttt{MaxCPUDelta},
\texttt{MaxMemoryDeltaMiB}, \texttt{MaxEvictions}), and an auto-allow
confidence threshold (0.85 validator default, 0.95 in the deployed
controller; actions below it are classified as
\texttt{require\_approval}); the 0.85 validator default is a
conservative operating default, not a threshold tuned on the evaluation
workloads, and all experiments still use approval-required execution. Second, \emph{contextual feasibility}:
validator-passing plans must satisfy pre-execution gates that are
deliberately outside the LLM: namespace ResourceQuota headroom for
replica and resource changes, PodDisruptionBudget feasibility before
voluntary eviction, a concurrent-intervention cap, and an optional UTC
maintenance window. A semantic guard is included for critical-path
contexts: when a \texttt{DiagnosisContext} contains a nonzero-duration
downstream \texttt{critical\_path} Deployment that differs from the
breached frontend, an agent-emitted \texttt{resize\_cpu} or
\texttt{scale\_out} must target that downstream Deployment or the
harness refuses to create the \texttt{RemediationPlan}. The guard does
not choose the action for the agent; it prevents semantically
inconsistent outputs from reaching the cluster. The guard runs as a
pre-submission check in the agent harness for agent-emitted plans;
deterministic-planner plans satisfy it by construction. Third, \emph{bounded
execution and recovery}: actions approved through the controller's
\texttt{RemediationPlan} approval path are patched to the declared
target only after those gates pass. For actions that mutate replica
count or container resource requests (\texttt{scale\_out},
\texttt{resize\_cpu}), the controller persists a pre-state snapshot to
the \texttt{RemediationPlan} status before the patch, and a rollback
monitor restores the pre-state if pod readiness deteriorates beyond a
configured grace period inside the observation window.

\begin{table}[t]
\centering
\small
\caption{Deterministic safety gates exercised by the validator and executor.}
\label{tab:safety-gates}
\begin{tabularx}{\columnwidth}{@{}p{0.40\columnwidth} X@{}}
\toprule
Gate & Expected decision \\
\midrule
Malformed JSON & \texttt{malformed} before validation \\
Unknown action & \texttt{deny} by action allow-list \\
Budget exceeded & \texttt{deny} by safety budget \\
Target mismatch & \texttt{deny} by semantic guard \\
Deny-listed target & \texttt{deny} by operator policy \\
Quota insufficient & \texttt{deny} before mutation \\
PDB blocks eviction & \texttt{deny} before eviction \\
Missing PDB & \texttt{deny} for voluntary eviction \\
Maintenance window & \texttt{blocked} before mutation \\
Concurrent cap & \texttt{blocked} before mutation \\
Valid approved action & executes bounded patch \\
\bottomrule
\end{tabularx}
\end{table}

These gates (Table~\ref{tab:safety-gates}) are exercised by replayed
model outputs and adversarial tests. Across the 78 recorded agent-replay outputs evaluated in
\S\ref{sec:agentic-ablation}, the validator allows 77 and denies 1
(over-budget \texttt{replicas{=}3} against
\texttt{MaxReplicasDelta{=}1}); every adversarial input reaches
\texttt{deny} or \texttt{malformed}. The evaluated controller uses an
approval-required \texttt{RemediationPlan} path: plans are created
pending approval, and the experiment harness approves validator-passing
plans for repeatability. Observe-only, canary-auto, and full-guarded
execution are compatible policy modes but are not enabled or evaluated
here; tenant hierarchy, traffic-level canaries, and aggregate SLO verdict
gates are validator extensions, not current experimental claims.

\section{Implementation and Scope}
\label{sec:implementation}

\sys{} is implemented as a Kubernetes-native controller with an
\otel-native data plane. Per-node DaemonSet collectors receive workload
and node signals; a gateway collector performs tail sampling and
\texttt{k8sattributes} enrichment; and a Go controller in the
\texttt{arbiter-system} namespace ingests OTLP/gRPC, maintains the
causal graph, builds \texttt{DiagnosisContext} objects, and emits
\texttt{RemediationPlan} custom resources (CRs). Each
\texttt{RemediationPlan} is a durable mutation envelope: it stores the
typed action, evidence, approval state, pre-state, phase, and observed
effect.

The controller follows the architecture of \S\ref{sec:overview}: it
retains graph state by age and span count, runs the deterministic
planner of \S\ref{sec:agents}, invokes the \texttt{SafetyValidator},
and executes only bounded Kubernetes actions. Along this path it
enforces ResourceQuota, PodDisruptionBudget, maintenance-window, and
concurrent-intervention gates; captures pre-state for reversible
actions; monitors rollback for \texttt{scale\_out} and
\texttt{resize\_cpu}; exposes SLO-sampler debug counters; and includes
the \texttt{critical\_path} planner branch of \S\ref{sec:agents}. The
live-gated agent harness remains outside the Go controller: it calls the
model on the same schema, then submits an agent-labeled
\texttt{RemediationPlan} through the same validator and approval path.

To avoid overstating autonomy or scale, the evaluation distinguishes
live evidence from artifact support. Live workload experiments cover
deployment-regression CPU-burn and pure-latency variants at $N{=}5$
each (\S\ref{sec:prelim-regression}), a DSB noisy-neighbor placement
matrix at $N{=}5$ (\S\ref{sec:noisy-placement}), an adversarial-input
sweep ($N{=}11$) at the validator, and injected-fault portability on
Online Boutique (\S\ref{sec:portability}). Agentic evidence consists of
a single-call replay ablation against three pinned Anthropic models plus
one live approval-gated execution path (\S\ref{sec:agentic-ablation}).
The artifact also contains critical-path traces, an alert-rollback
baseline, injected-fault harnesses, KWOK scale workflows,
kube-burner/ClusterLoader2 churn workflows, and a supporting DSB
CPU-pressure fallback-path run. That CPU-pressure run is fallback
evidence only because its SLO-sampler counters show no p99
observations in the pressure window (\S\ref{sec:cpu-pressure-supporting}).

At the planner boundary, \sys{} passes compact JSON contexts rather than
raw telemetry dumps. A diagnosis context carries six fields: SLO
symptom, request critical path, resource evidence, placement evidence,
recent changes, and the action vocabulary. When the graph holds a recent
root span, \texttt{critical\_path} attaches the chain from the slowest
root in the last 60\,s. The deterministic planner first checks
deployment-regression evidence, then uses \texttt{critical\_path} to
redirect resource repair to a dominant downstream Deployment when that
evidence disagrees with the breached frontend; only then does it fall
back to pod-CPU and node-pressure scoring. The replay corpus of
\S\ref{sec:agentic-ablation} focuses on deployment-regression action
selection and long-tail placement reasoning, while the live
critical-path experiments evaluate downstream target redirection
directly.
Listing~\ref{lst:context} shows the diagnosis context the controller
attaches to a \texttt{rollback\_canary} plan in the pure-latency
deployment-regression scenario of \S\ref{sec:prelim-regression}; the
\texttt{recent\_changes} field is what switches the action class from a
CPU-pressure response to \texttt{rollback\_canary}. The listing is a
regression case, so \texttt{critical\_path} is omitted there; it is
populated when a recent root span dominates the graph, as in the
downstream fault scenario of \S\ref{sec:critical-path-results}. The
listing predates the co-location enrichment: controller builds produced
after the evaluation campaign additionally populate
\texttt{placement\_evidence.colocated\_workloads} from the graph's
co-location edges, a planner-inert addition (\S\ref{sec:graph-plane})
absent from the captured artifact shown here.

\lstset{captionpos=b,abovecaptionskip=4pt}
\begin{lstlisting}[language=json,float=t,caption={Diagnosis context emitted on a deployment-regression breach.},label={lst:context}]
{
 "symptom": {
  "request_class": "nginx-web-server",
  "deployment":    "nginx-thrift",
  "hot_pod":       "nginx-thrift-...-dnztj",
  "slo":           "p99 < 300ms",
  "current_p99_ms": 504,
  "window":        "T-60s/60s"
 },
 "resource_evidence": [
  {"pod": "nginx-thrift-...-dnztj",
   "node": "agent-41", "cpu_milli": 503}
 ],
 "placement_evidence": [
  {"node": "agent-41",
   "node_cpu_percent":    9.58,
   "node_memory_percent": 9.26}
 ],
 "recent_changes": [
  {"kind": "Deployment", "name": "nginx-thrift",
   "event": "image :xenial -> :slow-latency-v2",
   "age_seconds": 15},
  {"kind": "Deployment", "name": "nginx-thrift",
   "event": "rollout gen 771 -> 772",
   "age_seconds": 15}
 ],
 "allowed_actions": ["deschedule_one", "resize_cpu",
                     "rollback_canary", "scale_out"]
}
\end{lstlisting}

\section{Evaluation Methodology}
\label{sec:eval}

\paragraph*{Questions} The evaluation is organized around five
questions. Q1 asks whether \sys{} selects the correct remediation action
on deployment regressions that resource-only autoscaling mishandles.
Q2 asks whether critical-path evidence redirects remediation from a
breached frontend to the responsible downstream Deployment. Q3 asks
whether an agentic planner on the same typed-action interface matches
deterministic behavior on covered contexts and extends it on uncovered
ones. Q4 asks whether deterministic safety gates contain malformed,
out-of-vocabulary, out-of-budget, or semantically inconsistent agent
outputs. Q5 asks whether the substrate transfers beyond a single
benchmark and separates live-runtime evidence from control-plane scale.

\begin{figure}[t]
\centering
\includegraphics[width=\columnwidth]{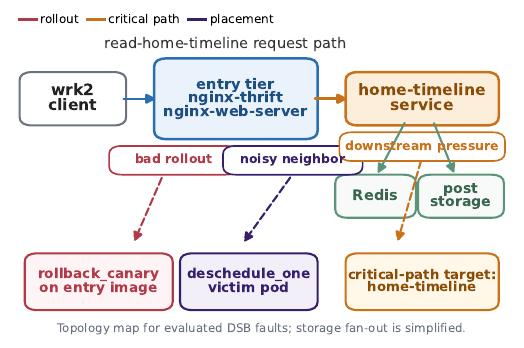}
\caption{Evaluated DeathStarBench Social Network request path and fault
map. The full benchmark contains additional services; the figure shows
the read-home-timeline path and the three injected fault sites used by
the regression, critical-path, and placement experiments.}
\label{fig:dsb-fault-map}
\end{figure}

\paragraph*{Benchmarks and load} DeathStarBench Social Network is the
primary live-workload benchmark because its request fan-out exposes
rollback, critical-path, and placement cases across roughly 25 services.
Google Online Boutique (\texttt{microservices-demo} v0.10.2, 12
HTTP/gRPC services) is used as a portability check rather than as the
source of comparative claims. DSB is driven by \texttt{wrk2} at
400\,RPS unless a scenario states otherwise; OB's bundled Locust-based
load generator is scaled to 3$\times$ its default replica count, at 50
users per replica, while the twelve application services keep their
default single-replica topology. Figure~\ref{fig:dsb-fault-map}
maps the DSB path exercised by the three live fault classes, so the
scenario descriptions use service names and remediation targets
consistently.

\paragraph*{Evidence tracks} Deployment regression is the main
SLO-control track for Q1. It has two flavors: a CPU-burn image that
tempts HPA to scale the bad code path, and a pure-latency image that
adds \texttt{ngx.sleep} without increasing CPU. Critical-path targeting
addresses Q2 by injecting pressure at
\texttt{home-timeline-service} while the SLO breach is observed at the
\texttt{nginx-thrift}/\texttt{nginx-web-server} entry path. Placement
repair evaluates a live noisy-neighbor case in which \sys{} must create
or reuse a spare, respect PDB feasibility, and evict one victim pod.
Agentic replay and one live-gated Sonnet path address Q3--Q4 on the
same \texttt{DiagnosisContext} and validator used by the controller.
Online Boutique and the scale/churn track address Q5. A CPU-pressure
run is reported separately as fallback-path evidence because its debug
counters show no p99 observations in the pressure window
(\S\ref{sec:cpu-pressure-supporting}).

\paragraph*{Baselines and metrics} The baselines mirror the
decision under test. Deployment regression asks whether a bad image is
rolled back rather than scaled (\texttt{none}, HPA, deterministic
\sys{}). Critical-path targeting asks whether trace evidence redirects
intervention from the breached frontend to the responsible downstream
Deployment (HPA/resource-only, deterministic \sys{}, live-gated Sonnet).
Because Q2 is a target-selection question, this track omits a
no-controller condition: without a controller there is no proposed
target to judge, so the informative contrast is resource-only versus
graph-aware planning.
Placement repair asks whether a controller can select and evict a victim
safely (\texttt{none}, HPA, topology/placement baseline, resource-only
\sys{}, placement-enabled \sys{}). We therefore emphasize action
correctness, target correctness, rollback rate, achieved RPS, mitigation
time, and validator decision. Mitigation time denotes the wall-clock
interval from fault injection to controller-confirmed completion of the
chosen remediation action, including pod-readiness verification for
mutating actions; the same definition is applied across flavors. Latency
distributions are reported when informative, but aggregate \texttt{wrk2}
p99 is not the lead metric for deployment regression: under slow-pod
throughput collapse, most completed requests remain baseline-fast and
the damaged window can appear above p99.9. DSB live comparisons use
\(N{=}5\) per condition unless noted; Online Boutique uses \(N{=}3\),
adversarial safety uses \(N{=}11\), and agentic replay uses all captured
contexts.

\subsection{Scale and Churn Track}
\label{sec:scale-churn}

The live DSB and OB experiments measure microservice runtime behavior
on the 4-node cluster; the scale track asks a narrower control-plane
question. Can the scheduler, API server, context generator, validator,
and replay path remain measurable as Kubernetes object cardinality and
churn increase? To answer this without pretending to run 10k real
microservice nodes, \sys{} separates two stresses. KWOK expands the
scheduler/API-server object space with 1k, 5k, and 10k-node targets: the
scheduler and API server remain real, while nodes and kubelets are
stubbed~\cite{kwok}; the 10k-node run did not complete within the
environment's resources and is not reported. Quantitative scale claims
are limited to completed runs and report scheduling throughput,
pending-pod latency, API-server
pressure, \sys{} context-generation overhead, and validator/replay
latency. On the physical cluster, kube-burner and ClusterLoader2 add
moderate pod and rollout churn under real API-server load~\cite{kubeburner,
clusterloader2}. This separation lets the paper claim control-plane
scaling evidence without converting fake-node experiments into
microservice-runtime claims.

\subsection{Artifacts and Reproducibility}
\label{sec:artifacts}

The reproducibility package is organized as an audit trail rather than
as an unstructured log dump. The article reports claim-level evidence:
architecture and safety substrate, deployment-regression recovery,
critical-path target redirection, noisy-neighbor placement repair,
Online Boutique portability, live-gated agent execution, and compact
scale evidence. The supplement binds those claims to replayable
artifacts: selected run manifests, \texttt{RemediationPlan} CRs,
controller log slices, resource samples, per-replicate \texttt{wrk2}
histograms, SLO-sampler debug counters, DSB-RegCtx contexts, prompts,
raw model outputs, validator decisions, adversarial inputs, KWOK/churn
logs, and scripts that regenerate every figure. This division keeps the
article within the page budget while preserving traceability from each
aggregate result to an executable harness or captured context.

\section{Evaluation Results}
\label{sec:results}

The evaluation is organized by claim rather than by implementation
component. The first experiments ask whether \sys{} chooses the right
Kubernetes action for a live DeathStarBench Social Network SLO breach;
the next experiments ask whether trace-derived context changes the
remediation target and whether guarded placement repair executes
safely. We then isolate the agentic interface, portability to Online
Boutique, and KWOK control-plane scale. This ordering keeps runtime
claims grounded in the 4-node live cluster while keeping offline replay
and fake-node scale evidence in their proper scope.

\subsection{Deployment-regression scenario}
\label{sec:prelim-regression}

This subsection answers Q1 by making action-class selection the
end-to-end decision. The fault is a bad image rollout on \texttt{nginx-thrift},
introduced without co-located noisy neighbors or pre-existing node
pressure. A controller that only sees resource pressure may add
capacity; the correct SLO remediation is instead to recognize the
recent change and roll back the regressed image. We instantiate this
fault in two complementary ways. In the CPU-burn flavor, the slow image
embeds a \texttt{stress-ng} worker, so HPA sees elevated CPU and the
question is whether the controller distinguishes rollback from resource
growth. In the pure-latency flavor, the slow image inserts
\texttt{ngx.sleep} in the read handler, adding latency without CPU work;
the question is whether the controller can act when the standard CPU
trigger is absent.

All conditions see the same demand and fault schedule. We compare
\texttt{none}, HPA (50\,\% CPU target on \texttt{nginx-thrift}, max 5
replicas), and deterministic \sys{} with the action vocabulary of
Table~\ref{tab:actions}. \texttt{wrk2} drives 360\,s of read traffic at
400\,RPS; the image changes at $T{+}60$\,s, and a safety-net restore is
scheduled at $T{+}300$\,s. Each condition/flavor pair is replicated
five times, with \texttt{RemediationPlan} CRs, controller logs,
resource samples, and deployment state archived for replay.

\begin{figure}[t]
\centering
\includegraphics[width=\columnwidth]{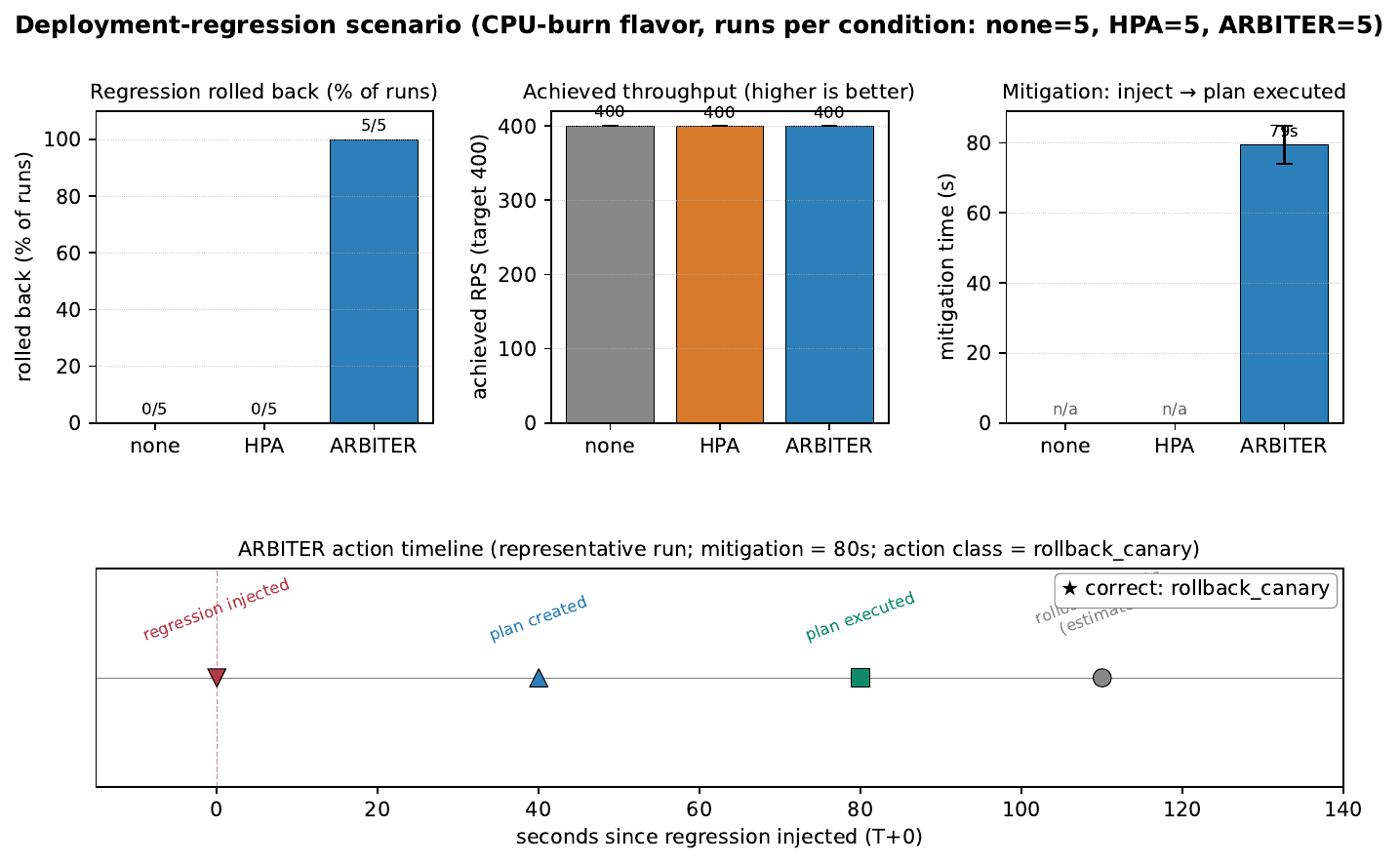}
\caption{Deployment-regression scenario, CPU-burn flavor at \(N{=}5\)
per condition. Top: rolled-back rate, achieved RPS, and mitigation
time. Bottom: representative \sys{} timeline; the \(\bigstar\) marks
the labeled-correct \texttt{rollback\_canary} action.}
\label{fig:regression-marquee}
\end{figure}

\begin{figure}[t]
\centering
\includegraphics[width=\columnwidth]{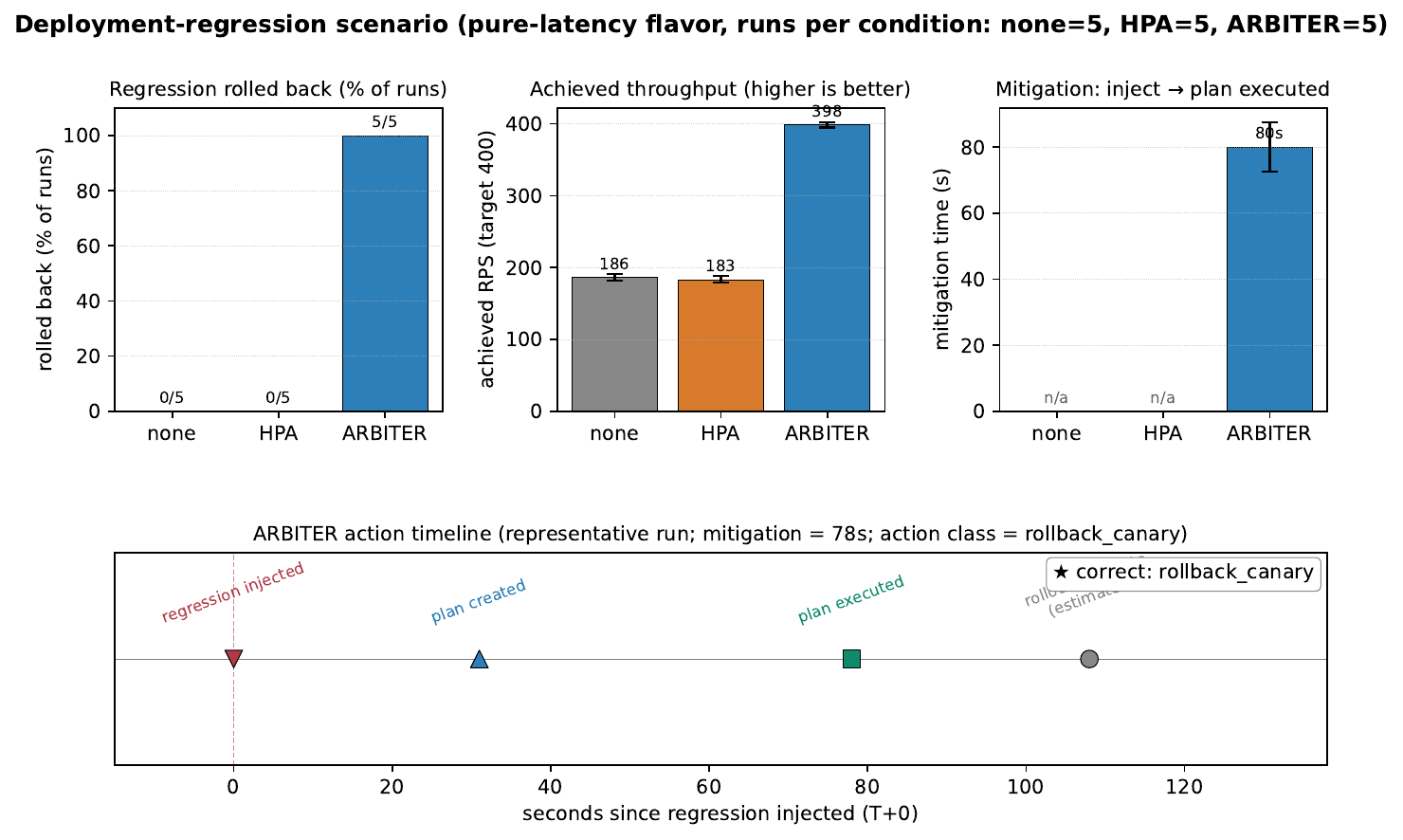}
\caption{Deployment-regression scenario, pure-latency flavor at
\(N{=}5\) per condition. Same panels as
Fig.~\ref{fig:regression-marquee}; \sys{} rolls back 5/5 runs,
mitigates in 80.0$\pm$7.5\,s, and sustains 398.3$\pm$3.5 RPS.}
\label{fig:regression-marquee-pl}
\end{figure}

\paragraph*{Action-class outcomes} Figures~\ref{fig:regression-marquee}
and~\ref{fig:regression-marquee-pl} show that deterministic \sys{}
selects \texttt{rollback\_canary} in all ten regression runs, across
five CPU-burn replicates and five pure-latency replicates. Each executed
plan records a direction-aware \texttt{observedEffect}. The executor uses
the \texttt{RecentChanges} event to roll \texttt{Deployment/nginx-thrift}
from the slow image variant back to the known \texttt{...:xenial} image,
rather than relying on an ambiguous previous \texttt{ReplicaSet} revision.
This rollback target is the core result of the scenario. In the CPU-burn
flavor, HPA chooses the wrong remediation action; every run scales to
the configured maximum. In the pure-latency flavor, HPA is effectively blind because
\texttt{ngx.sleep} yields the worker and leaves CPU near baseline. Thus
the same structured evidence handles both resource-only failure modes,
one caused by a misleading CPU signal and the other by no CPU signal at
all.

\paragraph*{Primary metrics} In the CPU-burn flavor, \sys{} mitigates in
79.4$\pm$5.4\,s while sustaining 399.8$\pm$0.1\,RPS. HPA also
sustains 399.8$\pm$0.1\,RPS in the CPU-burn flavor by scaling
\texttt{nginx-thrift} to five replicas in every run, but the added
replicas execute the same regressed image and therefore do not remediate
the rollout fault. In the pure-latency flavor, \sys{} mitigates in
80.0$\pm$7.5\,s and sustains 398.3$\pm$3.5\,RPS, while HPA reaches only
183.2$\pm$4.6\,RPS because it never executes a remediation. These
numbers use the deterministic planner; \S\ref{sec:agentic-ablation}
replays the captured
\texttt{DiagnosisContext} stream through the single-call agentic
planner. For experimental hygiene, the controller is restarted between
replicates to clear stale change-history events, and rollback execution
uses direction-aware image matching in every run.

\subsection{Critical-path downstream targeting}
\label{sec:critical-path-results}

This subsection answers Q2 by isolating a failure mode in which the SLO
breach is observed at the entry service while the remediable bottleneck
is downstream. The harness injects
CPU pressure on the node running \texttt{home-timeline-service}, while
the breached request class enters through
\texttt{nginx-thrift}/\texttt{nginx-web-server}. This setup creates a
controlled separation between symptom location and remediation target.
HPA and a resource-only controller operate on the exposed frontend
workload, so their available actions remain scale or resize decisions on
that workload. \sys{} instead follows the \texttt{critical\_path}
evidence in the \texttt{DiagnosisContext}, resolves downstream spans to
Kubernetes ownership metadata, and targets
\texttt{Deployment/home-timeline-service}.

\begin{figure*}[t]
\centering
\includegraphics[width=\textwidth]{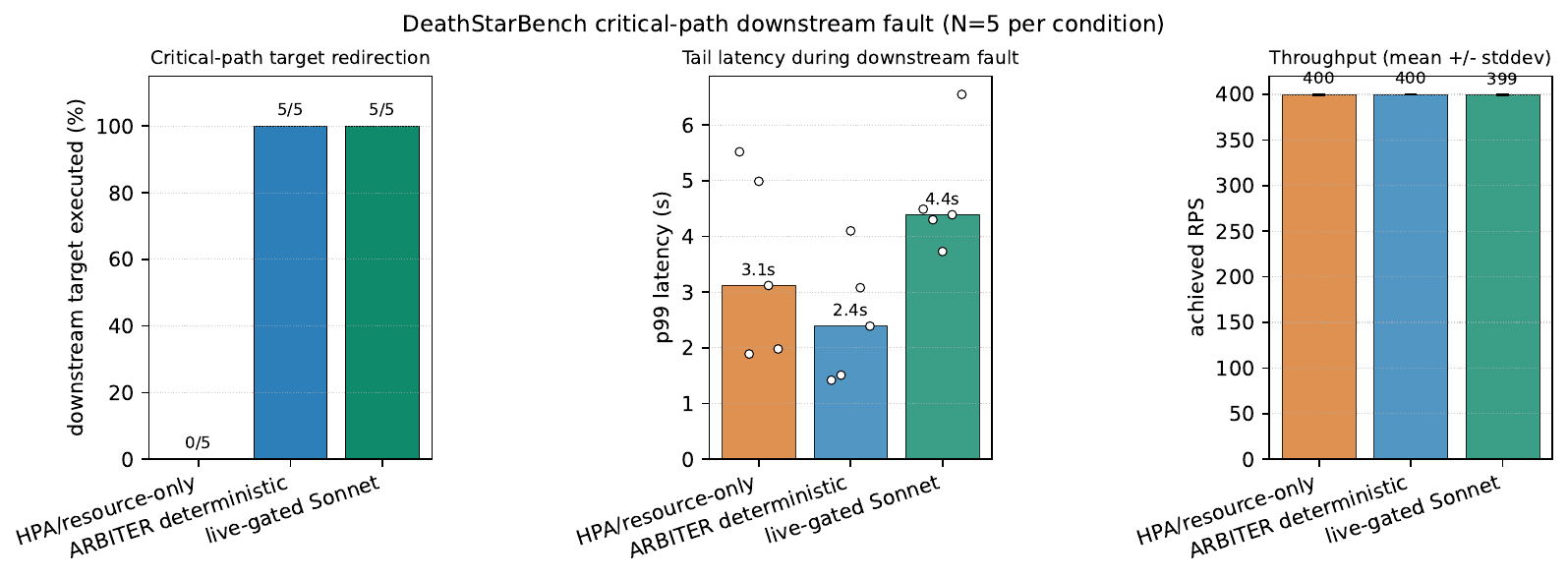}
\caption{DeathStarBench critical-path downstream fault. The breached
request enters through \texttt{nginx-thrift}, but the graph identifies
\texttt{home-timeline-service} as the remediable downstream target.
HPA/resource-only has no trace-derived target, while deterministic
\sys{} and the live-gated agent path execute through the same typed
validator and bounded executor.}
\label{fig:critical-path}
\end{figure*}

\paragraph*{Live-agent execution} The live-agent condition exercises
agentic planning at the same safety boundary as deterministic \sys{}. A
separate harness takes the controller-emitted \texttt{DiagnosisContext},
asks Sonnet for one finite-vocabulary typed-action JSON proposal, and
then hands that proposal to the deterministic substrate. The agent does
not hold execution authority; the harness creates an approval-gated
\texttt{RemediationPlan} only after the same schema validator and
critical-path semantic guard accept the action. Execution then uses the
unchanged bounded executor; this is the live Sonnet result in
Fig.~\ref{fig:critical-path}. Section~\ref{sec:agentic-ablation}
separately evaluates additional pinned models by replaying captured
contexts without further cluster mutation.

\paragraph*{Target-selection outcome} Figure~\ref{fig:critical-path}
summarizes the contrast. HPA/resource-only executes no downstream
remediation in five replicates. Deterministic \sys{} executes the
\texttt{home-timeline-service} action in every replicate (5/5), with
median p99 2.39\,s versus 3.12\,s for HPA at the same 399.7\,RPS
workload rate. The live-gated Sonnet harness also executes the
downstream target in every replicate (5/5), with median p99 4.39\,s at
399.5\,RPS. The Sonnet condition's larger tail is consistent with
inserting a single model call before plan materialization; the per-call
Sonnet latency in replay is 4.2$\pm$0.5\,s. Accordingly, the
live-Sonnet entry is reported as evidence of target correctness across
planners rather than latency competitiveness with the deterministic
path.

For this scenario, the primary metric is target correctness rather than
tail-latency dominance. The client observes the SLO violation at the
frontend, but the injected fault is on the node hosting
\texttt{home-timeline-service}. A
resource-only controller has no evidence that points to that downstream
Deployment, so it can only act on the exposed frontend workload. \sys{}
succeeds because the \otel-native \texttt{DiagnosisContext} preserves the
critical-path Deployment identity and presents it to the typed action
interface. The live-gated Sonnet run exercises the same boundary. The
model proposes a typed action, the semantic guard requires the downstream
target when critical-path evidence is present, and only an approved plan
reaches the bounded executor.

\subsection{Noisy-neighbor placement repair}
\label{sec:noisy-placement}

\paragraph*{Placement experiment} Placement repair asks whether
\sys{} can correct harmful live co-location rather than treating a hot
node only as a capacity problem. Each run starts with one clean
\texttt{nginx-thrift} replica, pins the load generator away from the
target node, allows HPA to scale to at most six replicas, and then injects
a high-CPU noisy neighbor on the selected \texttt{nginx-thrift} node. The
HPA baseline uses the same HPA maximum, PDB, pressure pod, and workload
rate as \sys{}. This scenario drives \texttt{wrk2} at an 800\,RPS
open-loop target rate, higher than the 400\,RPS used for the regression
and critical-path scenarios (\S\ref{sec:eval}), to keep the frontend
saturated through the pressure window. The \sys{} condition keeps HPA
but adds two-phase repair: wait for a ready spare and one allowed PDB
disruption, then evict one selected victim through the Kubernetes
eviction API.

\begin{figure*}[t]
\centering
\includegraphics[width=\textwidth]{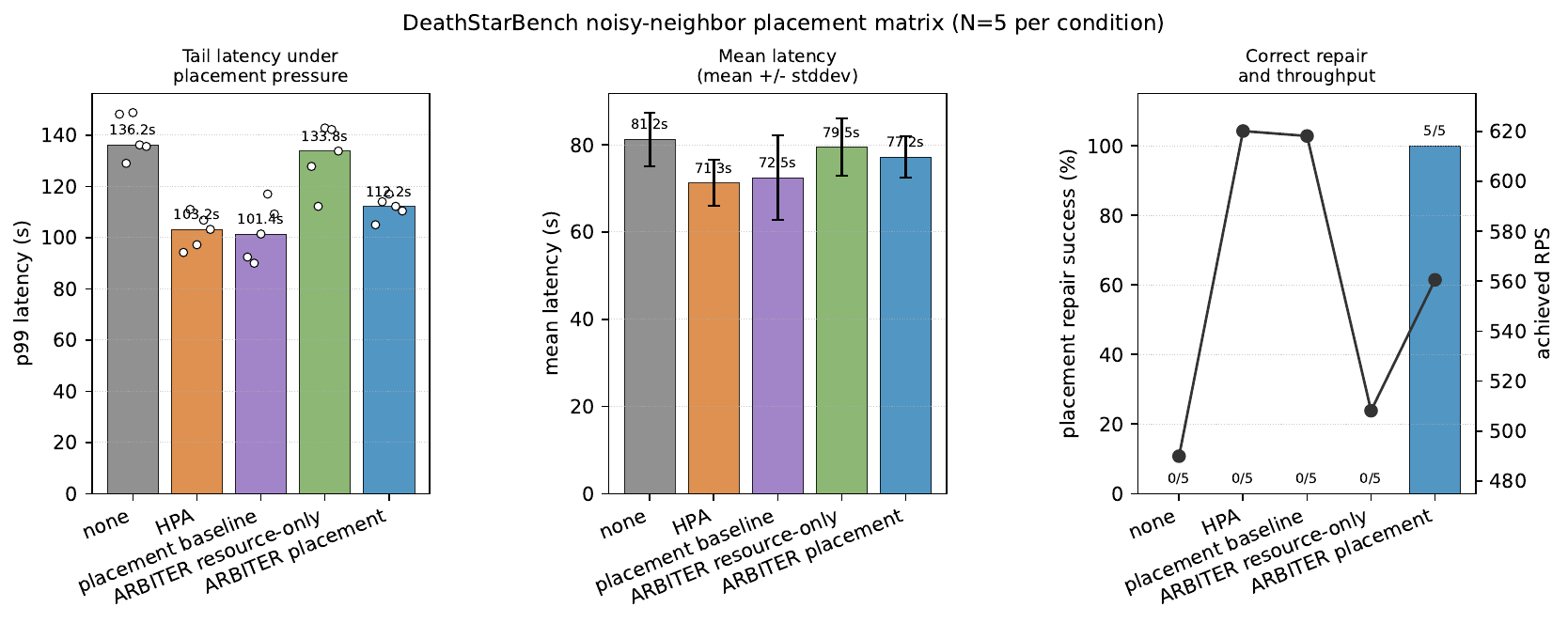}
\caption{DeathStarBench noisy-neighbor placement repair, \(N{=}5\) per
condition. Aggregate p99 is close because all conditions include the
pressure window. Repair success counts only PDB-gated victim evictions
completed through the protected two-phase path.}
\label{fig:noisy-placement}
\end{figure*}

\paragraph*{Repair outcome} Figure~\ref{fig:noisy-placement}
summarizes the placement result. Latency and throughput are secondary
because every condition includes the pressure window and aggregate p99
mixes clean and affected replicas. Median p99 is 136.2\,s for no
controller, 103.2\,s for HPA, 101.4\,s for the placement baseline,
133.8\,s for resource-only \sys{}, and 112.2\,s for placement-enabled
\sys{}. Throughput is also mixed: 620.1\,RPS achieved for HPA and
560.6\,RPS achieved for placement-enabled \sys{} against the 800\,RPS
offered load. We therefore make no latency-dominance claim.

The key result is guarded repair completion. Placement-enabled \sys{}
executes the protected two-phase path in all five runs; each
\texttt{RemediationPlan} records the hot pod, ready spare, and victim
eviction. HPA and the topology-spread baseline can add capacity or
influence initial placement, but they do not express live
diagnosis-and-repair as a typed, validator-gated eviction.

\subsection{CPU-pressure fallback}
\label{sec:cpu-pressure-supporting}

This auxiliary run bounds behavior when the SLO sampler is silent. Under
the same DSB workload, a 12-worker busy-loop pod is co-located with
\texttt{nginx-thrift}. The archived debug counters record zero p99
observations during the 240\,s pressure window, so the trigger is
hot-pod CPU evidence rather than a sampled SLO breach. \sys{} emits
\texttt{resize\_cpu} 16\,s after injection; p99 is 49.5\,s, compared
with 69.0\,s under no controller and 103.8\,s under HPA. We treat this
as executor/fallback validation; the primary SLO-remediation claim comes
from the p99-sampled deployment-regression and critical-path runs.

\subsection{Agentic-planning ablation}
\label{sec:agentic-ablation}

This ablation answers Q3 and Q4 by asking a narrow systems question:
what changes when the
planner behind the typed-action interface is replaced by a single-call
LLM planner? It is not a model ranking or an autonomous-operator claim.
The live Sonnet path in \S\ref{sec:critical-path-results} exercises
cluster execution; this subsection replays captured contexts to isolate
planner substitution, action validity, and validator behavior.

\paragraph*{Replay corpus} Each deployment-regression
\texttt{RemediationPlan} stores its \texttt{spec.diagnosisContext}
verbatim. We package ten such contexts as \textbf{DSB-RegCtx}: five
CPU-burn and five pure-latency regressions, all previously handled by
the deterministic planner. We add one noisy-neighbor context outside
the deterministic branch set and eleven adversarial inputs for the
validator. The replay uses three pinned models --- Sonnet 4.6
(\texttt{claude-sonnet-4-6}), Haiku 4.5
(\texttt{claude-haiku-4-5-20251001}), and Opus 4.7
(\texttt{claude-opus-4-7}) --- and translates every response into the
same \texttt{Action} struct used by the controller before invoking
\texttt{actions.SafetyValidator.DecisionFor}.

\paragraph*{Covered-context agreement} On the ten captured
regression contexts, the three models match deterministic \sys{} on all
thirty model-context pairs. Each emits \texttt{rollback\_canary} and
extracts the same target image,
\texttt{docker.io/yg397/openresty-thrift:xenial}, from
\texttt{recent\_changes}. This agreement is important because it shows
planner substitutability on a scenario already covered by rules: the
agentic planner can occupy the same typed interface without changing
the executor contract. Per-call latency is 2.3$\pm$0.3\,s for Haiku,
4.2$\pm$0.5\,s for Sonnet, and 3.7$\pm$1.1\,s for Opus; per-call cost
is \$0.0023, \$0.0074, and \$0.0410, respectively.

\paragraph*{Long-tail extension} The noisy-neighbor context
has no recent rollout event, so the regression branch does not fire.
The deterministic candidate scorer then prefers \texttt{resize\_cpu}
(score 0.83) over \texttt{deschedule\_one} (0.52) because the eviction
budget penalty dominates the confidence gap. That choice is intentionally
suboptimal in this context: adding CPU to a pod still co-located with a
saturating neighbor does not remove node-level interference. With the
complete action-semantics prompt, all three pinned models instead choose
\texttt{deschedule\_one} and cite
\texttt{placement\_evidence.colocated\_workloads}. The controller
populates \texttt{colocated\_workloads} from the graph's co-location
edges; the replayed long-tail context predates this and carries the
field hand-authored. The result is a
controlled example of the paper's thesis: structured agentic planning
can notice a long-tail field that the rule set does not consult, while
execution authority remains outside the model.

\paragraph*{Validator outcome} Across 78 recorded agent
responses (deployment-regression, long-tail, earlier prompt variants
used to calibrate the action menu, and
Online Boutique replay), every well-formed response reaches the
production validator: 77 are allowed and one is denied because
\texttt{replicas{=}3} exceeds \texttt{MaxReplicasDelta{=}1}. All eleven
adversarial cases are rejected. Unknown actions such as
\texttt{delete\_namespace} and \texttt{kubectl\_exec}, over-budget
parameters such as \texttt{replicas{=}99} and
\texttt{cpu\_request{=}50000m}, malformed JSON, missing required fields,
and empty inputs fail before execution. Thus the model can alter action
selection, but cannot bypass schema, policy, budget, or approval gates.

\paragraph*{Calibration lessons} The replay exposed two
data-shape issues rather than safety failures. An earlier prompt omitted
\texttt{deschedule\_one} from its prose action menu even though
\texttt{allowed\_actions} contained it; adding the missing semantics
moved all three models to placement repair without changing regression
answers. Some captured contexts also omitted an explicit namespace, so
models sometimes emitted \texttt{"default"} or \texttt{"arbiter"}; the
live executor ignores that string and uses controller configuration.

\subsection{Cross-microservice portability}
\label{sec:portability}

\paragraph*{Portability surface} This subsection answers the live-runtime
part of Q5. To check that the substrate is not DeathStarBench-specific,
we deployed Online Boutique
(\texttt{microservices-demo} v0.10.2), pointed its instrumented services
at the existing \sys{} \otel{} gateway, and retargeted only configuration:
the namespace, SLO definition, and allowed-action list. The causal graph
expanded to \texttt{services{=}12 pods{=}12} without Go-controller
changes, showing that service names, trace attributes, and Kubernetes
owners are carried by the \texttt{DiagnosisContext} schema rather than
by benchmark-specific code.

\begin{table}[t]
\centering
\caption{Online Boutique portability evidence.}
\label{tab:ob-portability}
\scriptsize
\begin{tabularx}{\columnwidth}{@{}p{0.21\columnwidth}X p{0.22\columnwidth}@{}}
\toprule
Check & Scope & Outcome \\
\midrule
Replay & 3 OB contexts \(\times\) 3 pinned models & 9/9 agree \\
Live fault & \(N{=}3\) injected noisy-neighbor runs & 3/3 execute \\
Action & frontend in every run; currencyservice in one run & CPU +100m (0.1 CPU) \\
\bottomrule
\end{tabularx}
\end{table}

\paragraph*{Portability outcome} Three natural Online Boutique
contexts (\texttt{frontend}, \texttt{currencyservice}, and
\texttt{loadgenerator}) replay with 9/9 action-class agreement across
the pinned models (Table~\ref{tab:ob-portability}). In the live
injected-fault run, all three replicates record nonzero frontend p99
samples and at least one breach, create approval-gated
\texttt{RemediationPlan} objects, and execute through the same bounded
controller path used by DSB. The executed action class is
\texttt{resize\_cpu}: every replicate increases
\texttt{Deployment/frontend} CPU request/limit by 100m (0.1 CPU) per container,
and one replicate also resizes \texttt{Deployment/currencyservice}
after an additional hot-service observation. We therefore treat Online
Boutique as portability evidence for the telemetry, typed-action, and
execution substrate; comparative performance claims remain on
DeathStarBench Social Network.

\subsection{Scheduler-scale evidence with KWOK}
\label{sec:kwok-results}

\paragraph*{Scale method} This subsection answers the control-plane
part of Q5. The live-workload results above are
separate from fake-node scale. In the KWOK track, Kubernetes scheduler
and API-server paths are real, while nodes and kubelets are simulated.
The harness creates fake Nodes and target Pods in retryable batches,
then archives convergence traces, Pod assignment snapshots, scheduler
logs, and API-server metrics. This measures scheduler/control-plane
behavior and \sys{} artifact handling under high object cardinality; it
does not emulate container startup, service traffic, or microservice
runtime interference.

\paragraph*{Scale outcome} The selected complete runs reach
1{,}000 fake nodes / 5{,}000 Pods and 5{,}000 fake nodes / 25{,}000
Pods. Both runs assign every target Pod (\(5{,}000/5{,}000\) and
\(25{,}000/25{,}000\)); unscheduled count is zero. The local scheduling
summary reports p99 pending-Pod latency of 1.0\,s for both scales, with
maximum observed assignment latency of 2.0\,s and 4.0\,s, respectively.
We therefore use KWOK to bound reproducible scheduler/control-plane
scale, while all microservice runtime claims remain on the live DSB and
Online Boutique experiments.

\section{Related Work}
\label{sec:related}

\sys{} draws on three lines of work. First, SLO-oriented microservice
controllers such as FIRM~\cite{firm}, Autothrottle~\cite{autothrottle},
Sinan~\cite{sinan}, GRAF~\cite{graf}, and AutoMan~\cite{automan} show
that dependency-aware resource control and learned policies can improve
microservice QoS. These
systems motivate critical-path and SLO feedback, but their action spaces
are primarily resource-control policies. \sys{} instead exposes a
Kubernetes remediation interface that includes rollback, placement
repair, resizing, and scaling behind the same typed safety boundary.
At the cluster-capacity layer, AgileSphere~\cite{agilesphere} combines
predictive and reactive node autoscaling with warm-pool promotion to
reduce capacity-pressure SLO violations; \sys{} is complementary,
selecting bounded remediation actions and targets for the workloads
running on whatever capacity such systems provision.

Second, observability and RCA systems such as Sage~\cite{sage},
Seer~\cite{gan2019seer}, MicroRCA~\cite{microrca}, and
Sieve~\cite{sieve} use logs, metrics, or traces to localize
microservice performance problems. \sys{} uses similar
evidence for localization, but treats localization as an intermediate
artifact: the output is a bounded \texttt{DiagnosisContext} that must be
translated into a finite, auditable Kubernetes action before any cluster
mutation occurs.

A newer line of LLM-for-operations work studies incident summarization,
root-cause explanation, mitigation recommendation, and autonomous
cloud-agent evaluation. LLM-based incident recommenders~\cite{ahmed-llm-incident},
RCACopilot~\cite{rcacopilot}, RCAgent~\cite{rcagent}, and
AIOpsLab~\cite{aiopslab} show that language models can synthesize
heterogeneous operational evidence and invoke tools over realistic
cloud tasks. \sys{} is complementary: it does not ask the model to
operate the cluster. The model, when used, only proposes a
finite-vocabulary typed action over a pre-correlated
\texttt{DiagnosisContext}; schema checks, semantic guards, quotas,
disruption feasibility, approval, execution, and rollback remain outside
the LLM.

Third, Borg~\cite{borg}, Omega~\cite{omega}, and Kubernetes'
cluster-control lineage~\cite{borgomega} establish the substrate on
which modern schedulers build. Recent surveys organize custom
Kubernetes scheduling~\cite{carrion-k8s-scheduling},
extension-oriented scheduling~\cite{rejiba-custom-scheduling}, and
scheduler algorithms~\cite{senjab-k8s-scheduling}; complementary
systems study network-aware placement~\cite{marchese-network},
communication-aware microservice scheduling~\cite{marchese2022communication},
MOTAS~\cite{motas}, and DRS~\cite{drs}. Kubernetes itself exposes the
scheduling framework~\cite{k8ssched}, KEDA~\cite{keda},
descheduler-style eviction~\cite{k8sdescheduler}, and topology-spread
primitives~\cite{k8stopologyspread}.
\sys{} does not replace these mechanisms; it composes them under an
SLO-oriented, \otel-native evidence layer and a shared deterministic
validator. Its contribution is the substrate boundary: swappable
reasoning (\texttt{DiagnosisContext} in, finite-vocabulary action out)
with deterministic validation and bounded execution for both
deterministic and agentic planners.

\section{Claim Boundaries}
\label{sec:boundaries}

\sys{} bounds authority: planners may propose remediation only when an
SLO violation is explained by telemetry and maps to a finite Kubernetes
action; the cluster never receives unbounded model output. It is not an
application debugger;
correctness bugs, data corruption, and external dependency outages may
still require developer intervention, and diagnosis quality depends on
telemetry coverage. To make that boundary observable, the implementation
records SLO-sampler counters for observed root spans, matched and
dropped request classes, sample counts, and breach decisions. The
agentic layer is evaluated as typed action selection, not autonomous
operation: schemas, action vocabularies, deny-lists, budgets,
quota/PDB, maintenance and concurrency gates, approval, pre-state
capture, and rollback remain deterministic.

\section{Conclusion}
\label{sec:conclusion}

\sys{} demonstrates that SLO-oriented Kubernetes remediation can use
agentic planning without making the model the operator. OpenTelemetry
and Kubernetes state become a bounded \texttt{DiagnosisContext};
planners choose from a finite action vocabulary; and policy, budgets,
approval, execution, and rollback remain deterministic. In live
DeathStarBench faults and Online Boutique portability runs, the same
boundary supports rollback, critical-path targeting, placement repair,
and portable resource repair. The calibrated
thesis the paper substantiates is two-way: \emph{the deterministic
substrate is what makes agentic planning safe; agentic planning is what
makes the deterministic substrate extensible.} Safe agentic cloud
control rests on explicit evidence and action boundaries around trusted
enforcement.

\bibliographystyle{IEEEtran}
\bibliography{references}

\end{document}